\documentclass[]{elsart}

\usepackage{epsfig}
\usepackage{graphics}
\usepackage{graphicx}
\usepackage{amsmath}
\usepackage{url}
\usepackage{subfigure}

\newtheorem{finding}{Finding}
\newtheorem{hypothesis}{Hypothesis}

\begin{document}

\begin{frontmatter}

\title{How $\beta$-skeletons lose their edges}

\author{Andrew Adamatzky}

\address{University of the West of England, Bristol, United Kingdom\\ andrew.adamatzky@uwe.ac.uk}

\date{\today}

\begin{abstract}

\vspace{0.5cm}

\noindent
A $\beta$-skeleton is a proximity graphs with node neighbourhood defined by continuous-valued parameter $\beta$.
Two nodes in a $\beta$-skeleton are connected by an edge if their lune-based neighbourhood contains no 
other nodes. With increase of $\beta$ some edges a skeleton are disappear. We study how a number of edges in 
$\beta$-skeleton depends on $\beta$. We speculate how this dependence can be used to discriminate between 
random and non-random planar sets. We also analyse stability of $\beta$-skeletons and their sensitivity to perturbations. 

\end{abstract}

\maketitle

\end{frontmatter}

\section{Introduction}

A planar graph consists of nodes which are points of Euclidean plane and edges which are straight segments connecting the points. A planar proximity graph is a planar graph where two points are connected by an edge if they are close in some sense. Usually a pair of points is assigned certain neighbourhood, and points of the pair are connected by an edge if their neighbourhood is empty.  Delaunay triangulation~\cite{delauanay}, relative neighbourhood graph~\cite{jaromczyk} and Gabriel graph~\cite{matula_1980}, and indeed spanning tree, are most known examples of proximity graphs.  

$\beta$-skeletons~\cite{kirkpatrick} is  a unique family of proximity graphs monotonously parameterised by $\beta$. Two neighbouring points of a planar set are connected by an edge in $\beta$-skeleton if a lune-shaped domain between the points contains no other 
points of the planar set. Size and shape of the lune is governed by $\beta$.

\begin{figure}[!tbp]
\centering
\includegraphics[scale=1]{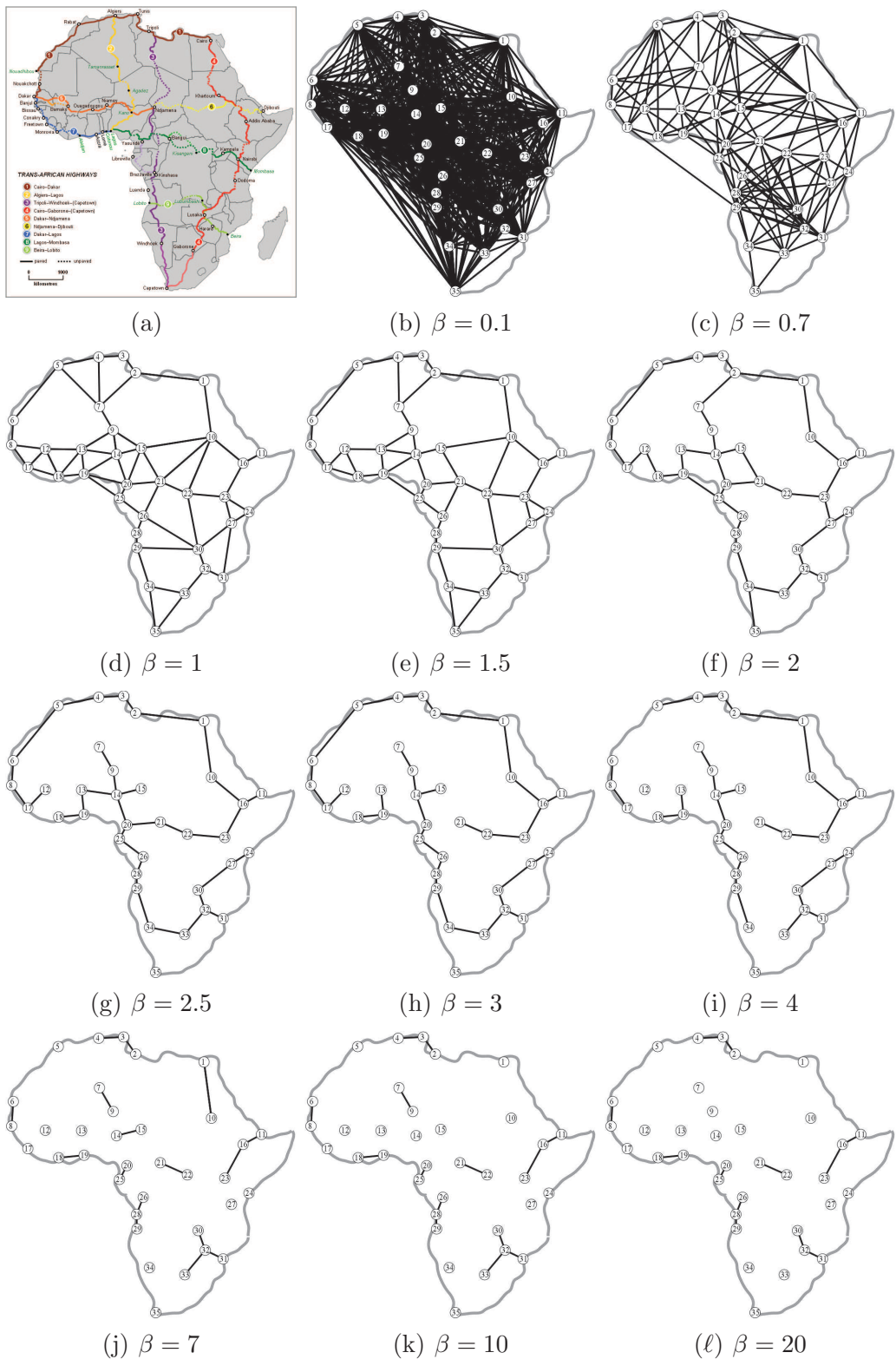}
\caption{$\beta$-skeleton approximation of African highways.
(a)~original scheme of trans-African highways~\cite{parry}. (b)--(l)~$\beta$-skeletons on major urban areas are
illustrated for selected values of $\beta$ from 0.1 to 20. See details in~\cite{adamatzky_bioevaluation,adamatzkykayem2013}.}
\label{africa}
\end{figure}

Why is it necessary to study properties of $\beta$-skeletons? The $\beta$-skeletons are eminent representatives of 
the family of proximity graphs. Proximity graphs  found their applications in fields of science and engineerings:  
image processing and computational morphology: e.g. curve reconstruction from a set of 
planar points~\cite{Amenta_1998}, approximation of road networks~\cite{watanabe_2005, watanabe_2008},
geographical variational analysis~\cite{gabriel_1969,matula_1980,sokal_2008}, 
evolutionary biology~\cite{magwene_2008}, 
spatial analysis in biology~\cite{legendre_1989,dale_2000,dale_2002,jombart_2008}, 
simulation of epidemics~\cite{toroczkai_2008}. Proximity graphs are used in physics 
to study percolation~\cite{billiot_2010} and analysis of magnetic field~\cite{sridharan_2010}. 
Engineering applications of proximity graphs are in message routing in ad hoc wireless networks, see e.g.~\cite{li_2004,song_2004,santi_2005,muhammad_2007,wan_2007}, and visualisation~\cite{runions_2005}. 
Road network analysis is yet another field where proximity graphs are invaluable. Road networks are well 
matched by relative neighborhbood graphs, see e.g. study of Tsukuba central 
district~\cite{watanabe_2005, watanabe_2008}. Biological transport networks also bear remarkable 
similarity to certain proximity graphs. Foraging trails of ants
and protoplasmic networks of slime mold \emph{Physarum polycephalum}~\cite{adamatzky_ppl_2008,adamatzky_bioevaluation}
are most striking examples. 

In our previous works on approximation of man-made road networks with slime mould and proximity graphs~\cite{adamatzky_bioevaluation}
we found that $\beta$-skeletons provides sufficiently good approximation of highway network in many countries for $\beta$ lying between 1 
and 2 (Fig.~\ref{africa}ab). A $\beta$-skeleton, in general case, becomes disconnected for $\beta>2$ and continues losing its edges with further increase of $\beta$  (Fig.~\ref{africa}c--l). Are sections of road networks, which survive longer with increasing $\beta$ bear any particular 
importance? We did find, see details in~\cite{adamatzky_bioevaluation,adamatzkykayem2013}, that by tuning value of $\beta$ we can, in principle, make a difference between paved and unpaved roads in Trans-African highway network, however an ideal matching between a $\beta$-skeleton 
and a high-way graph was every achieved. Thus we got engaged with studies of dynamics of $\beta$-skeletons. Some finding we made so far are outlined in present paper. We answer the following questions.  How a rate of edge disappearance depends on $\beta$? For what configurations 
of planar points  $\beta$-skeleton does not lose its edges with increase of $\beta$? Can we differentiate between random and non-random configurations of planar points by a curve of their $\beta$-driven edge disappearance?

\section{$\beta$-skeletons}
\label{themodel}

\begin{figure}[!tbp]
\centering
\includegraphics[scale=1]{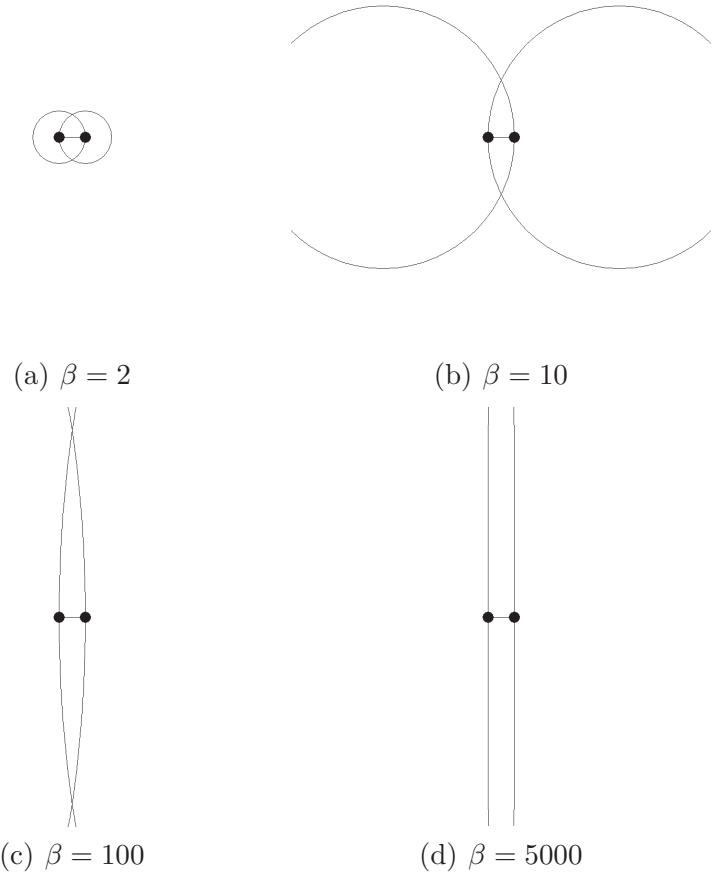}
\caption{Examples of lunes ($\beta$-neighbourhoods) of two planar points (small circles) for various values of $\beta$.}
\label{lunes}
\end{figure}

Given a set $\mathbf V$ of planar points, for any two points $p$ and $q$ we define 
$\beta$-neighbourhood $U_\beta(p,q)$ as an intersection of two discs 
with radius $\beta |p-q| / 2$ centered at points $((1-\frac{\beta}{2})p,\frac{\beta}{2}q)$ and 
$(\frac{\beta}{2}p, (1-\frac{\beta}{2})q)$,  $\beta \geq 1$~\cite{kirkpatrick,jaromczyk}, see examples of the lunes in Fig.~\ref{lunes}.
Points $p$ and $q$ are connected by an edge in $\beta$-skeleton if the pair's $\beta$-neighbourhood contains no 
other points from $\mathbf V$.

A $\beta$-skeleton is a graph $G_\beta({\mathbf V})= \langle {\mathbf V}, {\mathbf E}, \beta \rangle$, 
where nodes ${\mathbf V} \subset {\mathbf R}^2$, edges $\mathbf E$, and for $p, q \in {\mathbf V}$ 
edge $(pq) \in \mathbf E$ if $U_\beta(p,q) \cap {\mathbf V}/\{p, q\} = \emptyset$. Parameterisation 
$\beta$ is monotonous: if $\beta_1 > \beta_2$ then  $G_{\beta_1}({\mathbf V}) \subset  G_{\beta_2}({\mathbf V})$~\cite{kirkpatrick,jaromczyk}. 
A $\beta$-skeleton is Gabriel graph~\cite{matula_1980} for $\beta=1$ and the skeleton is relative neighbourhood graph for $\beta=2$~\cite{kirkpatrick,jaromczyk}.

\section{Edges losses in skeleton on random planar sets}

\begin{figure}[!tbp]
\centering
\includegraphics[scale=1]{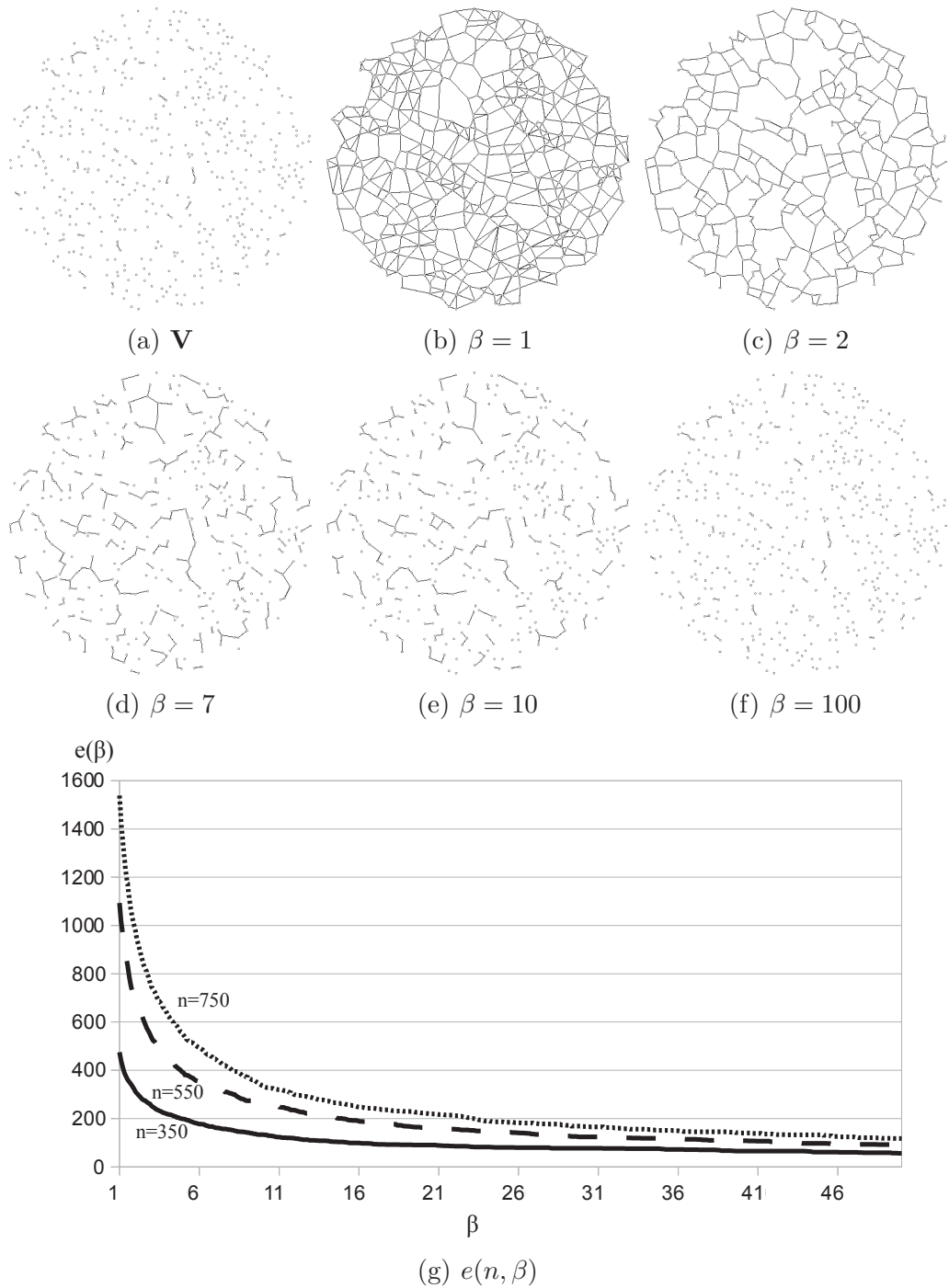}
\caption{Skeletons of a random planar set  $\mathbf{V}$ (a)  lose their edges with 
increase of $\beta$. (b--h)~Examples of $\beta$-skeletons on planar set
of 500 discs, radius 2.5 each,  randomly distributed in a disc radius 250.
(i)~Example power curves $e(n,\beta)$, $1 \leq \beta \leq 50$, $n=350$ (solid line),
$n=550$ (dashed line) and $n=750$ (fine dashed line), values of $\beta$ are incremented by 0.1.}
\label{randomdiscexamples}
\end{figure}

\begin{figure}[!tbp]
\centering
\includegraphics[scale=1]{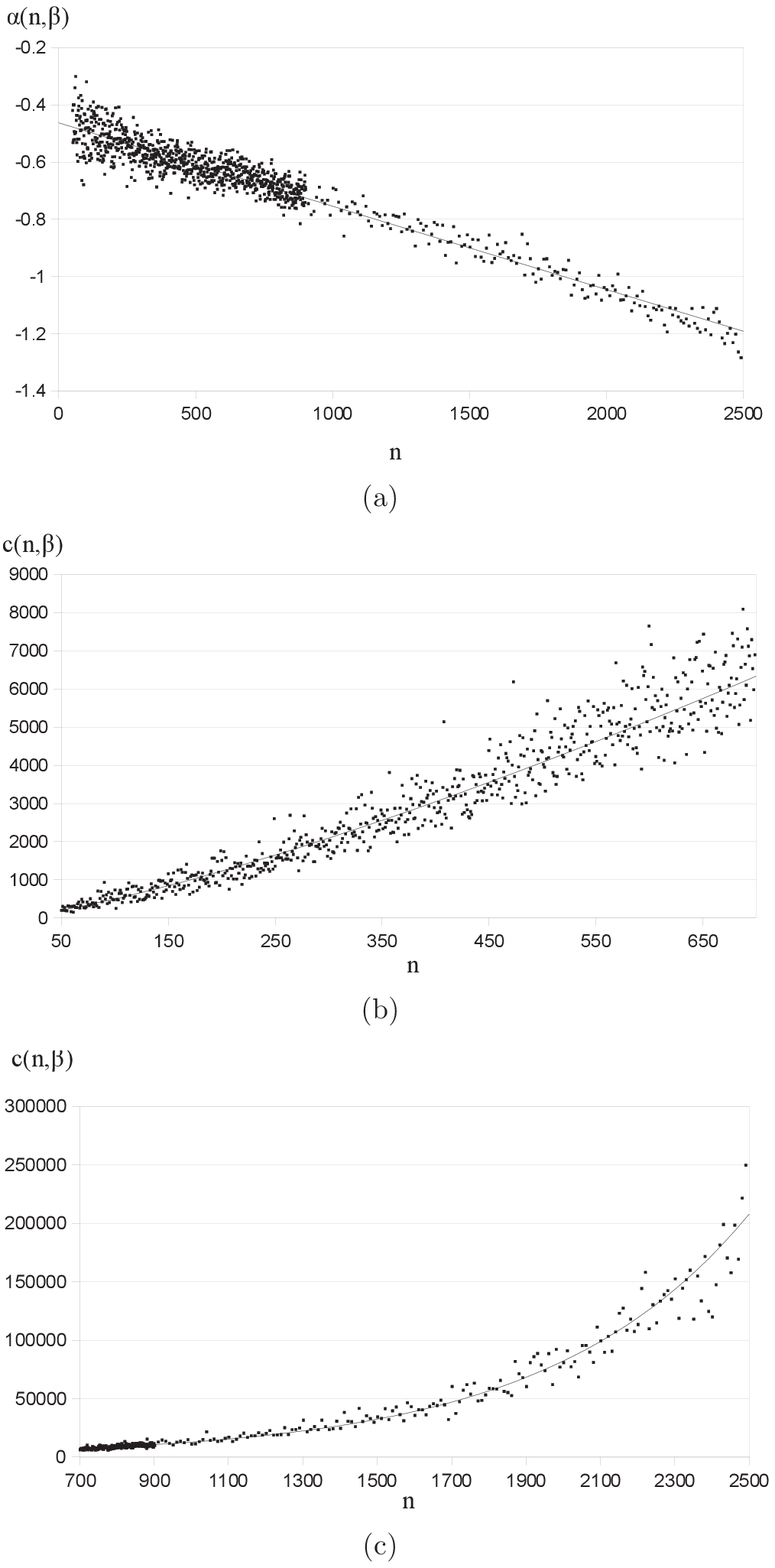}
\caption{Approximation of coefficients $c(n,\beta)$ and ${\alpha(n,\beta)}$ 
of $e(n,\beta)=c(n,\beta) \cdot n^{\alpha(n,\beta)}$ for $n$ up to 2500: 
(a)~$\alpha(n,\beta)$, linear approximation line $\alpha(n,\beta)=-0.0003 \beta \cdot 0.4628$, $R=0.925$; 
(b)~$c(n, \beta)$, $50 \leq n \leq 700$, $c(n,\beta)=1.2072 \cdot \beta^{1.3075}$, $R=0.957$;
(c)~$c(n, \beta)$, $700 \leq n \leq 2500$, $c(n,\beta)=1998.5 \cdot 1.0019^\beta$.}
\label{coefficients}
\end{figure}

To analyse rate of edge losses in $\beta$-skeletons of random planar sets we represented planar points
by $n$ discs,  centres of the discs form set $\mathbf V$. Each disc has a radius 2.5 units and 
the discs are randomly distributed in a large disc with radius 250  (Fig.~\ref{randomdiscexamples}).   
For $n$ up to 2500 and $\beta$ varying from 1 to 50 we calculated number of edges $e(n,\beta)$ 
in $\beta$-skeletons  (Fig.~\ref{randomdiscexamples}b--h). Example curves are shown in 
Fig.~\ref{randomdiscexamples}i.  Data points $e(n,\beta)$ are approximated by power curve 
$e(n,\beta)  \sim c(n,\beta) \cdot \beta^{\alpha(n, \beta)}$. 

\begin{finding}
$\beta$-skeletons of random planar sets lose their edges by power law. 
Number decreases proportionally to $\beta^{\alpha}$,  $\alpha<0$. Absolute value of 
$\alpha$ is linearly proportional to number of planar points in the sets. 
\end{finding}

To uncover how $c(n,\beta)$ and $\alpha(n, \beta)$ depends on $n$ we approximated 
$e(n,\beta)$ for planar sets $n=50, 60, 70 \ldots, 2500$ and $\beta=1,2,3, \ldots 50$. 
Data points calculated  are shown in Fig.~\ref{coefficients}. 

Coefficient $\alpha(n, \beta)$ 
linearly decreases (increases in its negative values) with increase of a number of nodes 
(Fig.~\ref{coefficients}a). Coefficient $c(n,\beta)$ increases proportionally to $\beta^b$ 
for $n \leq 700$ (Fig.~\ref{coefficients}b) and the coefficient grows proportionally to 
$d^\beta$ for $n > 700$ (Fig.~\ref{coefficients}c), where $1 \leq b, d \leq 2$.

\section{Differentiating between random and non-random sets}

In previous section we demonstrated that presence of even minor impurities in originally regular arrangement of planar points can be detected directly in the shape of edge disappearance curve $e(n, \beta)$. This leads us to the following hypothesis.

\begin{figure}[!tbp]
\centering
\includegraphics[scale=1]{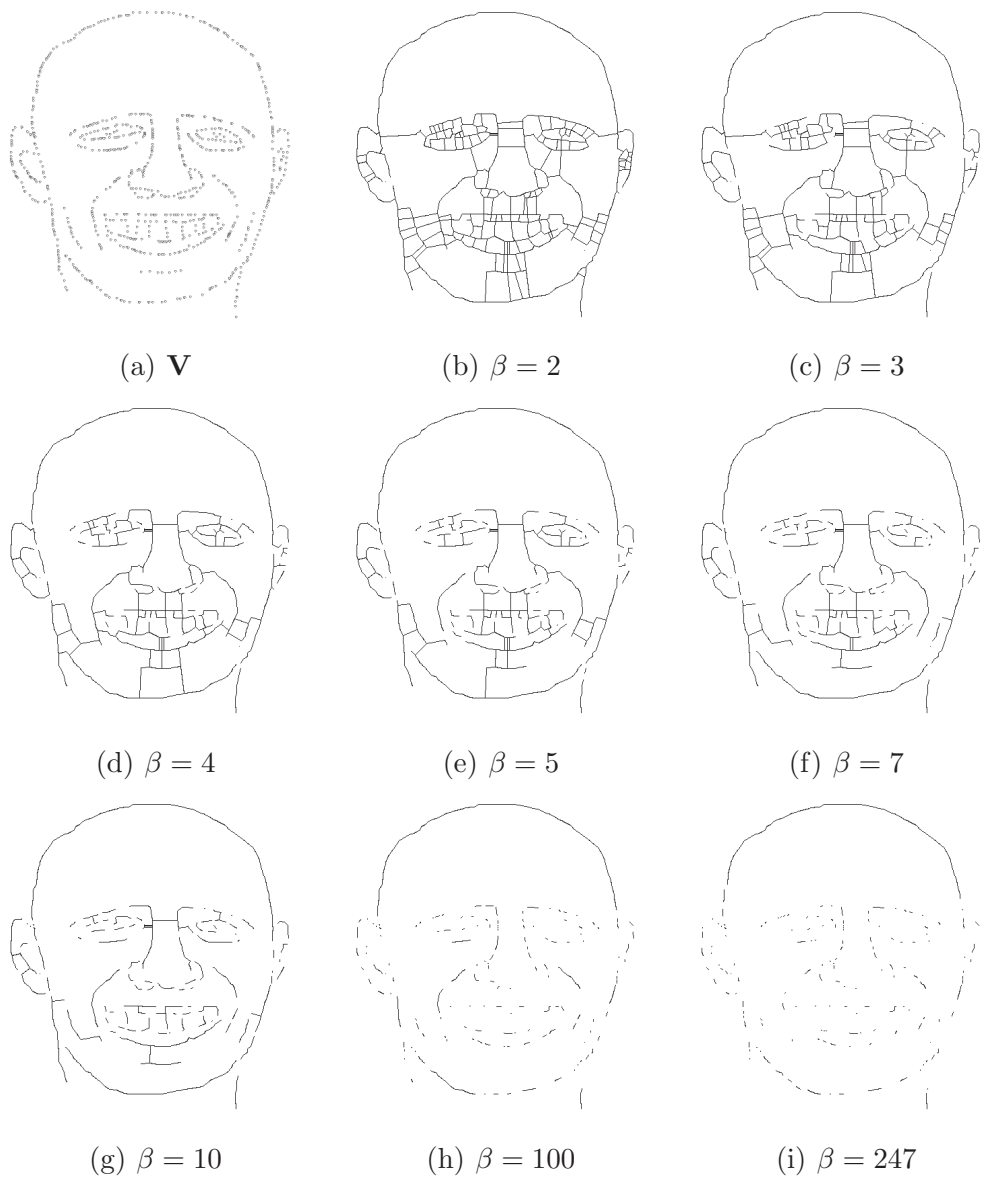}
\caption{A  drawing of a face   represented by planar points $\mathbf V$  with $n=823$ nodes 
and its  $\beta$-skeletons for $2 \leq \beta \leq 247$; $n=823$.}
\label{face}
\end{figure}

\begin{figure}[!tbp]
\centering
\includegraphics[scale=1]{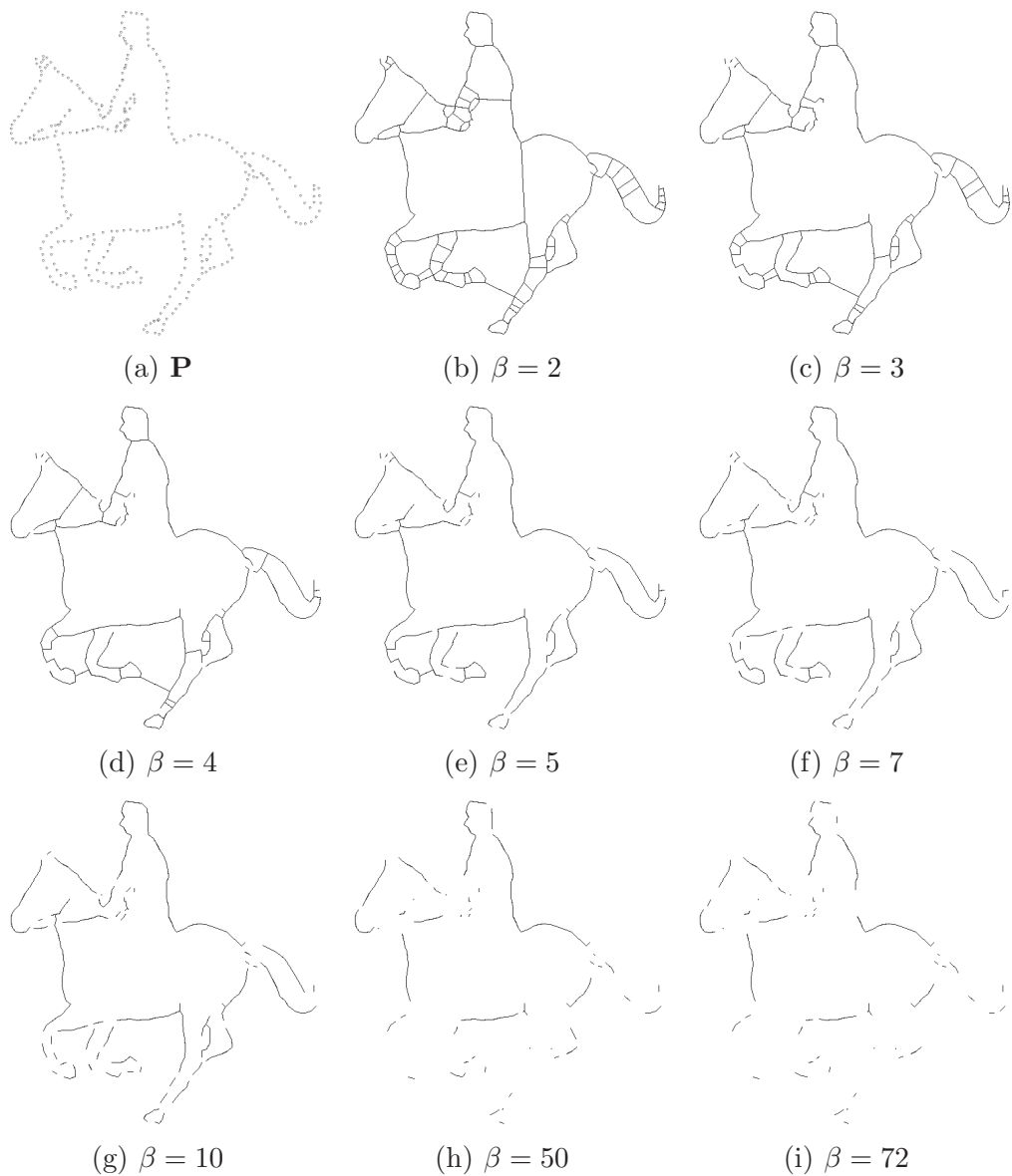}
\caption{A drawing of a horseman represented by planar points $\mathbf V$ with $n=351$ nodes  
and the $\beta$-skeletons of $\mathbf V$ for selected values of $2 \leq \beta \leq 72$.}
\label{horse}
\end{figure}

\begin{figure}[!tbp]
\centering
\includegraphics[width=0.9\textwidth]{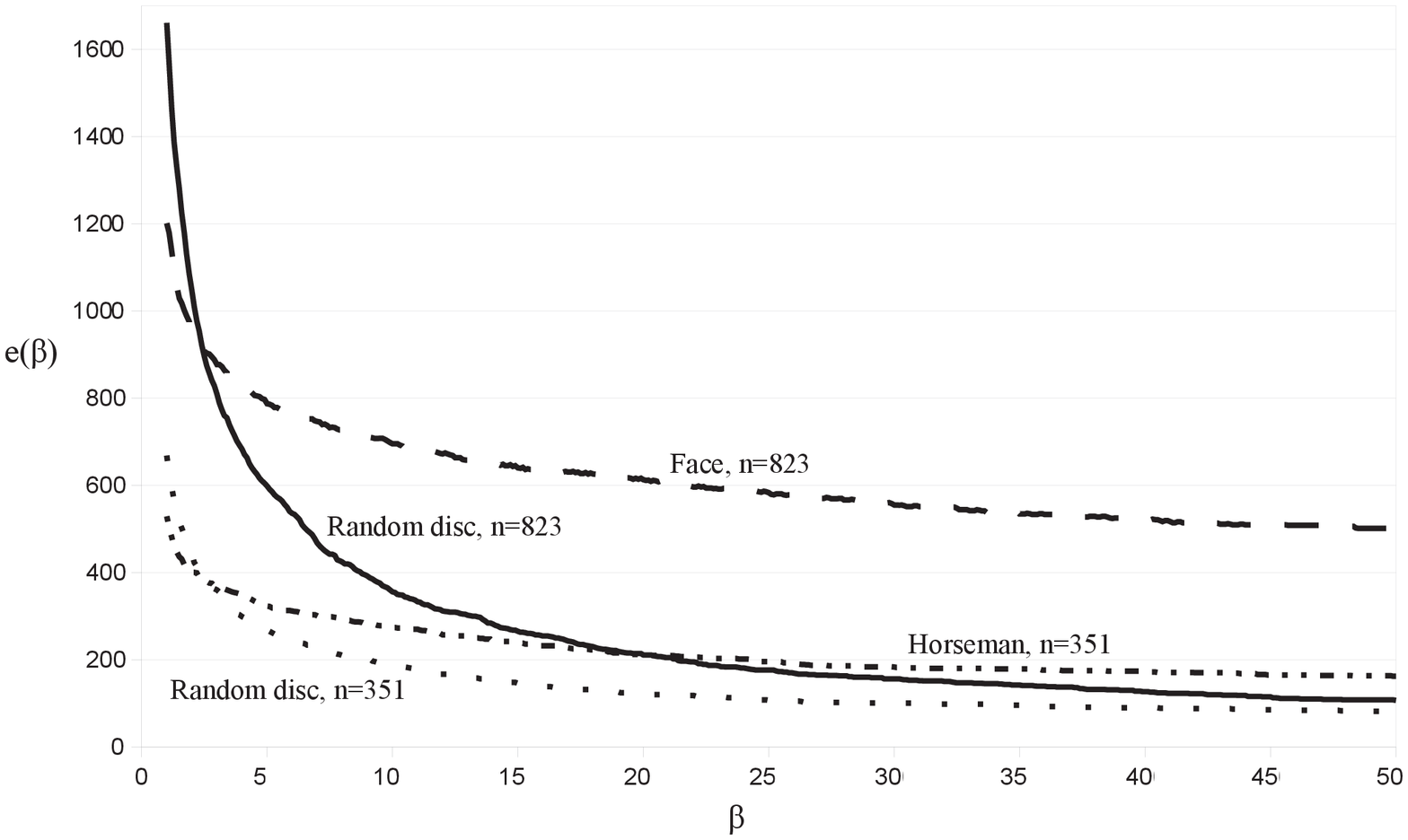}
\caption{Edge disappearance curves $e(n,\beta)$ of $\beta$-skeletons constructed over planar points representing face (Fig.~\ref{face}), 
shown by  dashed line; horseman (Fig.~\ref{horse}), shown by line of dashes and dots; and two random set of planar points, distributed in discs, 
with the same number of points as face and horseman, shown by solid and dotted lines, respectively. Values of $\beta$ are incremented by 0.1.}
\label{horse_face_graph}
\end{figure}

\begin{hypothesis}
Random planar sets can be differentiated from non-random sets by a shape of edge disappearance curve $e(n, \beta)$.  
\end{hypothesis}

We do not aim to prove the hypothesis in present paper but rather demonstrate its viability in two examples. We represented drawings of a face 
and a horseman and in sets of planar points (Figs.~\ref{face}a and~\ref{horse}a). Evolution of $\beta$-skeletons 
of these sets, associated with removal of certain edges of $\beta$-skeletons, leads to formation of contour like representations of 
the images (Figs.~\ref{face} and~\ref{horse}). We calculated edge disappearance curves $e(n, \beta)$ for fixed $n$ and $\beta$ changing 
from 1.0 to 50 with increment 0.1 (Fig.~\ref{horse_face_graph}, dashed line and dash-dots line). We also produced curves $e(n, \beta)$
for random sets of planar points, with the same numbers of points, distributed in a disc radius 250 units (Fig.~\ref{horse_face_graph}, 
solid line and dotted line). 

\begin{table}
\caption{Coefficients of power regression approximation $c(n,\beta) \cdot \beta^{\alpha(n, \beta)}$ of $\beta$-driven edge disappearing 
in $\beta$-skeletons of non-random and random planar sets.}
\begin{tabular}{lccc}
Planar set  & $n$   & $c(n, \beta)$ & $\alpha(n, \beta)$ \\ \hline
Face 		   &  823  & 1797.9       	& -0.296 \\
Random set	   & 823   & 10078.9 		& -0.73 \\
Horseman	   &  351  & 1077.8 	& -0.306 \\
Random set	   & 351   & 2159.0 	& -0.5344 \\
\end{tabular}
\label{facehorsetable}
\end{table}

The data are approximated by power regression $c(n,\beta) \cdot \beta^{\alpha(n, \beta)}$ with coefficients shown in 
Tab~\ref{facehorsetable}. The coefficients were calculated using a non-linear least square technique using Gauss-Newton 
algorithm~\cite{bjork_1996}.

Based on Fig.~\ref{horse_face_graph} and Tab.~\ref{facehorsetable} we can conclude that random planar sets
have initially higher number of edges than non-random sets however they exhibit higher rate of edge disappearance driven by $\beta$.  
For $\beta=1$ a number of edges in the skeleton of face is 0.72 of edges comparing to ta number of edges in a skeleton of a random planar set
with the same number of points; and skeleton of horseman has 0.79 of edges of its corresponding random set. The skeletons of non-random sets 
have almost the same number of edges as skeletons of random sets at $\beta=2.4$ (face) and $\beta=2.7$ (horseman). After that value of $\beta$ number of edges in skeletons of random sets decreases substantially quicker than number of edges of skeletons of non-random sets. 
Thus, at $\beta=50$ $\beta$-skeleton of face has 4.68 times more edges than a skeleton of its corresponding random set, and skeleton of horseman has 2 times more edges than skeleton of a random set. 

The two examples considered are not at all enough to make any rigorous conclusions, however we can speculate that the difference between random and non-random sets occurs when $\beta$ is changed from $2$ to $3$ (i.e. almost at the same time when skeletons are at first becoming disconnected); and, it is enough to compute $\beta$-skeletons till $\beta=10$ because for such value of $\beta$ number of edges in skeletons of non-random sets 1.5 times higher than a number of edges in skeletons of random sets.

\section{Stability and impurities}

Not all $\beta$-skeletons lose their edges with increase of $\beta$.  Special cases of stable $\beta$-skeletons 
are discussed in present section.  Let $\mathbf{B}_{ab}$ be an open half-plane bounded by an infinite straight 
line $l_a$ passing through $a$, perpendicular to segment $(a,b)$ and containing $b$; and  $\mathbf{B}_{ba}$ 
be an open half-plane bounded by an infinite straight line $l_b$ perpendicular to segment $(a,b)$, passing through 
$b$ and containing $a$.  Let $\mathbf{H}_{ab} =  \mathbf{B}_{ab} \cap \mathbf{B}_{ba}$.
When $\beta$ becomes extremely large, tends to infinity, a $\beta$-neigbourhood of any two neighbouring
points $a$ and $b$ tends to $\mathbf{H}_{ab}$.  A $\beta$-skeleton of planar set $\mathbf{V}$ is stable if for 
any $a, b \in \mathbf{V}$  $\mathbf{L}_{ab}$ does not contain any points from $\mathbf{V}$ apart of $a$ and $b$. A stable 
$\beta$-skeleton retains its edges for any value of $\beta>1$.

\begin{figure}[!tbp]
\centering
\includegraphics[scale=.5]{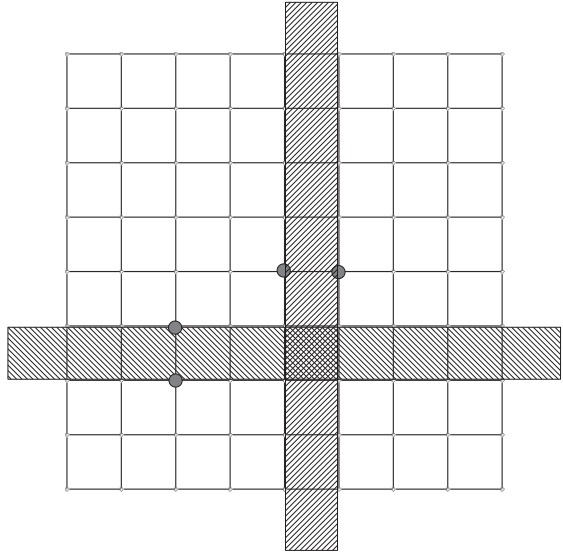}
\caption{Rectangular lattice is a stable $\beta$-skeleton. $\beta$-neighbourhoods, $\beta \rightarrow \infty$, of two pairs of nodes (marked by large discs) are shown by hatched areas. }
\label{rectangularlatticescheme}
\end{figure}

A most obvious example of a stable $\beta$-skeleton is a skeleton built on a set of planar points arranged in a 
rectangular array. The rectangular $\beta$-skeleton conserves its edges for any value of $\beta$ (Fig.~\ref{rectangularlatticescheme}).
The rectangular attice is stable because for any two neighbouring nodes $a$ and $b$ intersection of their 
half-planes $\mathbf{H}_{ab}$ fits between rows or columns of nodes without covering any nodes. 

\begin{figure}[!tbp]
\centering
\includegraphics[scale=1]{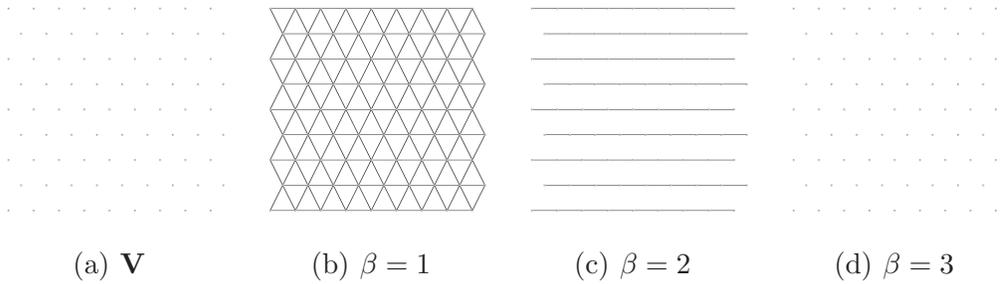}
\caption{Transformation of $\beta$-skeletons built on a hexagonal array of planar points $\mathbf{V}$. }
\label{hexagonal}
\end{figure}

\begin{finding}
Regularity does not guarantee  stability.  
\end{finding}

In Fig.~\ref{hexagonal} we show skeletons of a hexagonal arrangement of planar points (Fig.~\ref{hexagonal}a).  A skeleton is a 
hexagonal lattice for $\beta=1$  (Fig.~\ref{hexagonal}b). All diagonal edges of the lattice disappear when 
 $\beta=2$ (Fig.~\ref{hexagonal}c). With further increase of $\beta$ to 3 horizontal edges vanish  
 (Fig.~\ref{hexagonal}d)  and all nodes of the original planar set become isolated.

\begin{figure}[!tbp]
\centering
\includegraphics[scale=1]{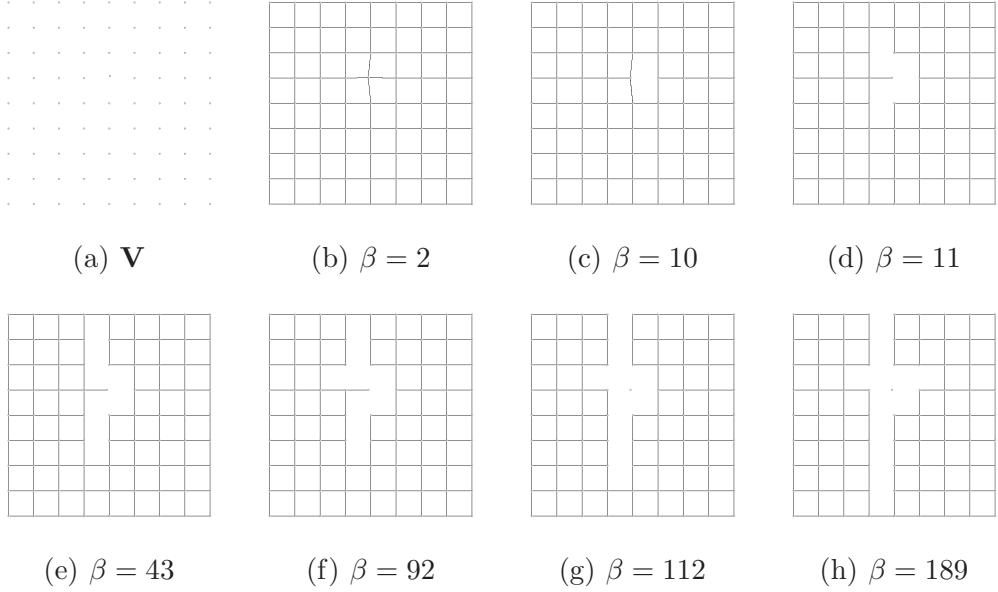}
\caption{Transformation of $\beta$-skeleton on planar points arranged in a rectangular array (a) 
with a single 'defective' node in  5th column and 4th row.}
\label{defectINorthogonal}
\end{figure}

\begin{finding}
Stable $\beta$-skeletons are sensitive to perturbations.
\end{finding}

Stable $\beta$-skeletons are sensitive to even slight distortions of a regular arrangement of elements $\mathbf{V}$. This is illustrated in Fig.~\ref{defectINorthogonal}. One node in the otherwise perfect uniform and regular rectangular array of planar points (Fig.~\ref{defectINorthogonal}a) gets 
its coordinates slightly randomised, so its $x$ coordinate is different from other nodes in its row, and its $y$ coordinate is different from other nodes in its column.  A localised distortion of the skeleton can be seen in edges linking node in 5th column and 4th row with its four neighbours (Fig.~\ref{defectINorthogonal}b). With increase of $\beta$ the 'defective' node starts losing its edges (Fig.~\ref{defectINorthogonal}c). With further increase of $\beta$ the defect induced edge elimination propagates along row and columns adjacent to the defective node (Fig.~\ref{defectINorthogonal}d--g). Eventually a value of $\beta$ reached where no more edges are removed and the skeleton remains stable under subsequent growth of $\beta$ (Fig.~\ref{defectINorthogonal}h).

\begin{figure}[!tbp]
\centering
\includegraphics[scale=1]{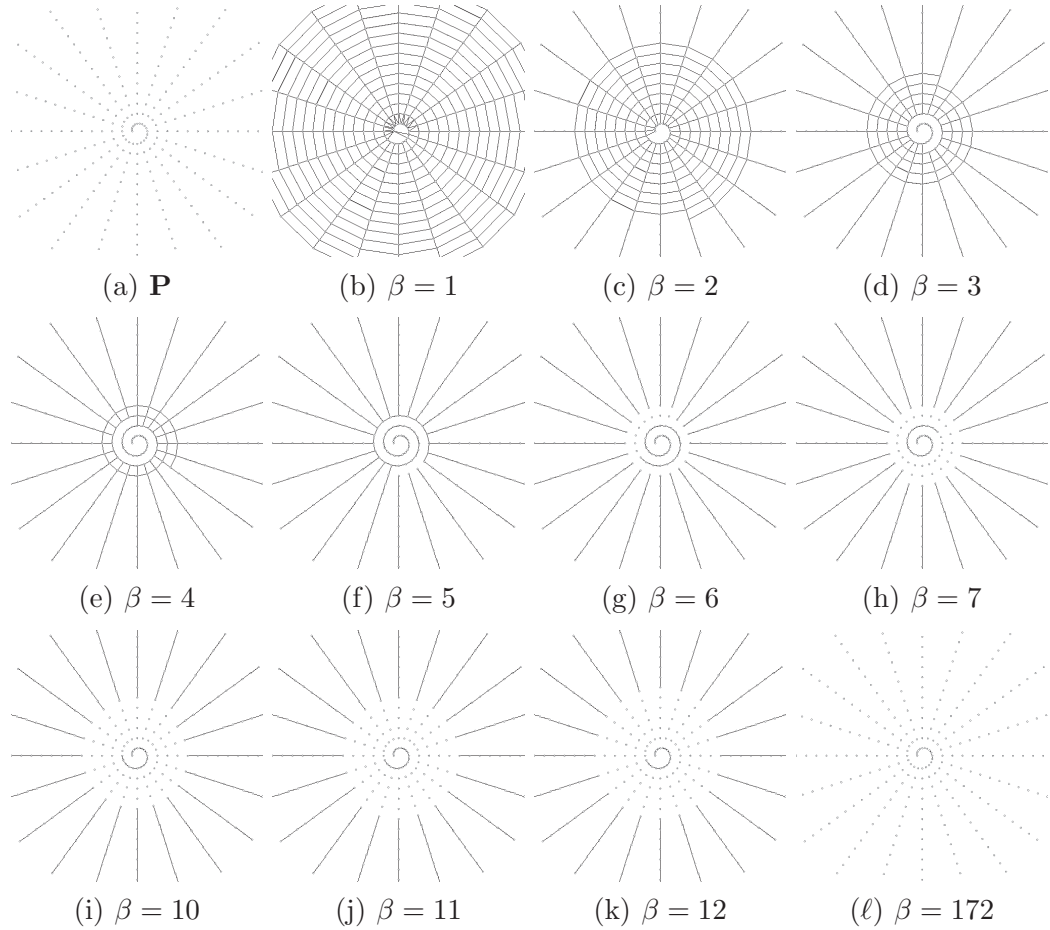}
\caption{Restructuring of a spider-web $\beta$-skeleton by increasing $\beta$.}
\label{spiderweb}
\end{figure}

When regular $\beta$-skeletons are 'dissolved' by increasing $\beta$ order of an edge disappearance is 
determined by  the edge location. In Fig.~\ref{spiderweb}a we consider a planar set which core nodes are arranged 
in a spiral and other nodes lined up in rays.   The planar set is spanned by spider-web looking $\beta$-skeleton for 
$\beta=1$  (Fig.~\ref{spiderweb}b).  The spiral part of the skeleton retracts back towards its centre when
$\beta$ increases from 1 to 4 (Fig.~\ref{spiderweb}cde). At the value $\beta=5$ only nodes which where originally in the spiral shape (two rotations) 
and nodes aligned in rays are connected by edges of the $\beta$-skeleton (Fig.~\ref{spiderweb}f). Further increase of $\beta$ causes retraction of the original spiral and dilution of rays, with edges disappearing centrifugally (Fig.~\ref{spiderweb}g--l).

\begin{figure}[!tbp]
\centering
\includegraphics[scale=1]{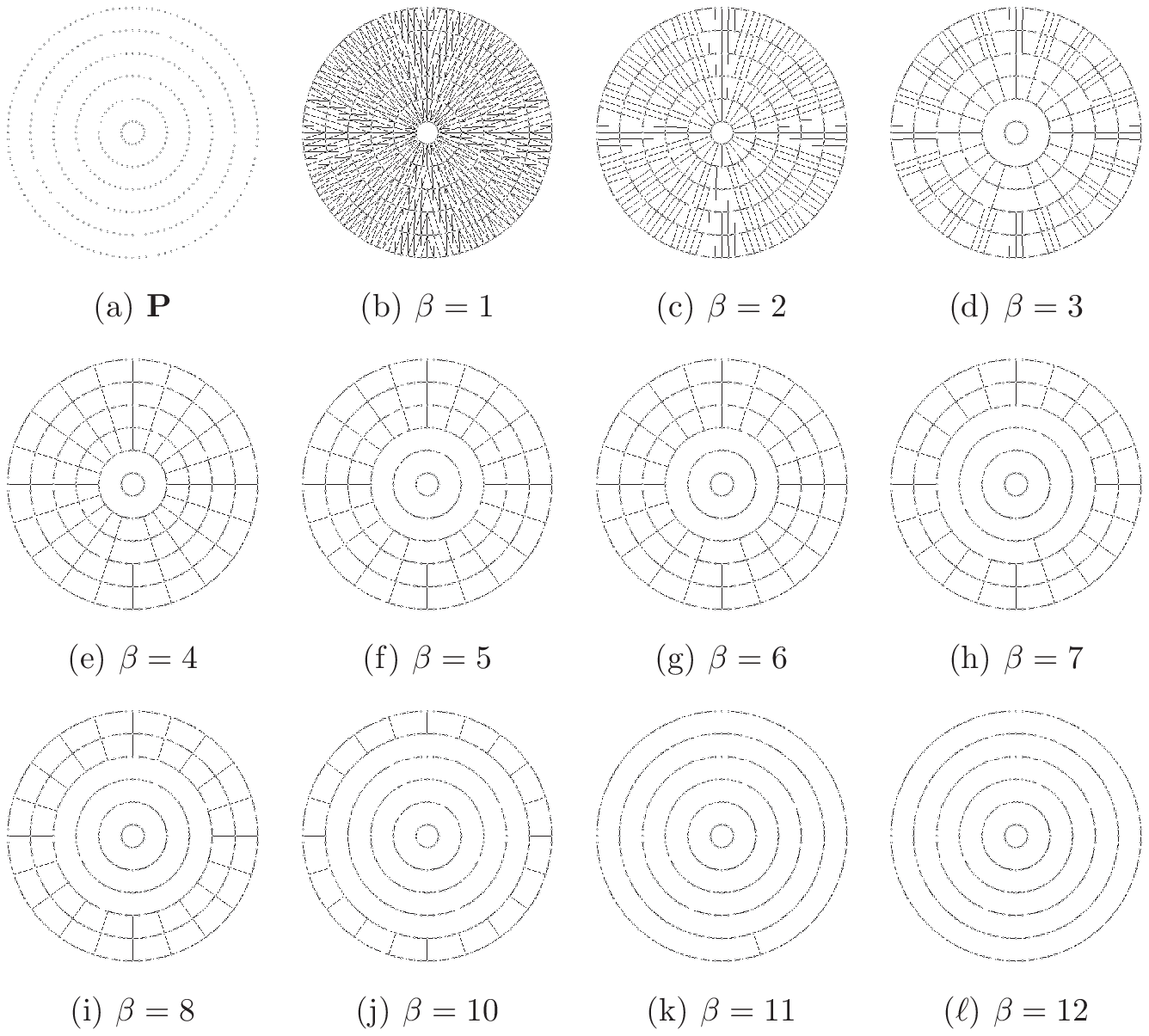}
\caption{A $\beta$-skeletons of planar points arranged in six nested circles centred at the same point $c$.}
\label{circles}
\end{figure}

\begin{figure}[!tbp]
\centering
\includegraphics[scale=1]{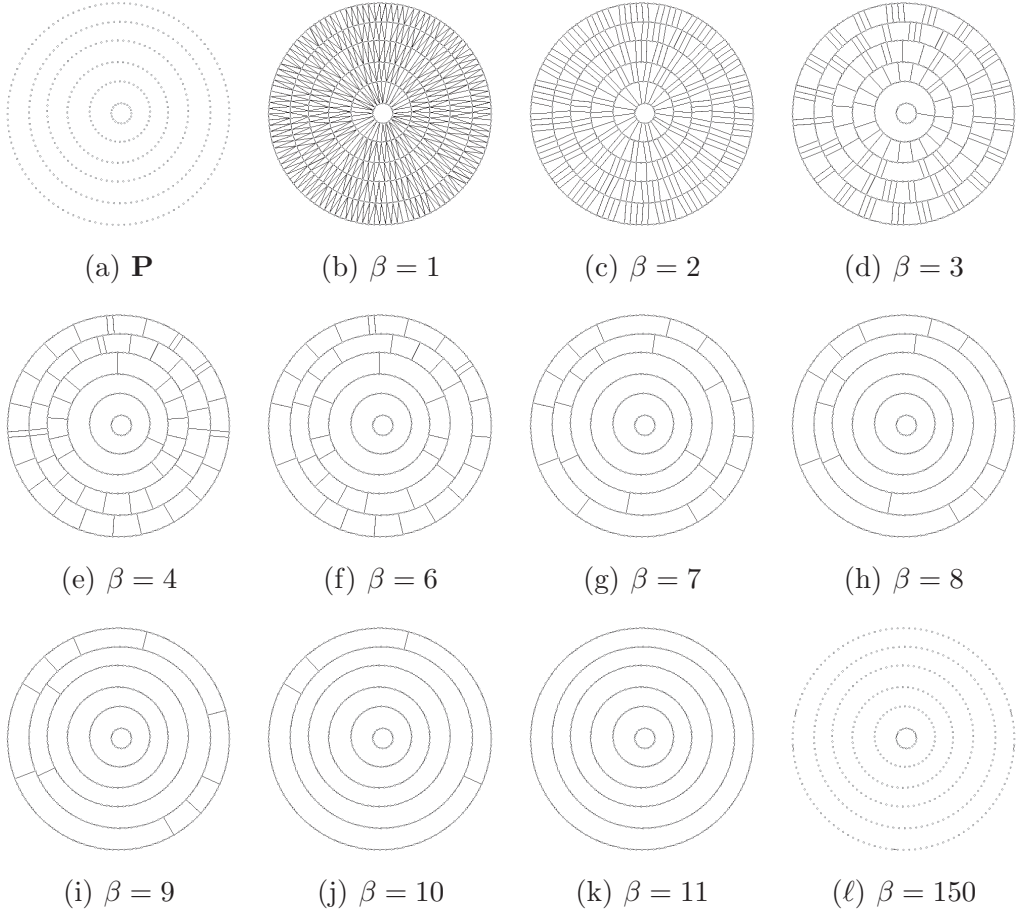}
\caption{$\beta$-skeletons of planar points arranged in six nested circles centred with minor, $[-5,5]$ units, 
random deviations from original point $c$ ($c$ is used as centre in Fig.~\ref{circles}).}
\label{circlesrandomcentres}
\end{figure}

\begin{figure}[!tbp]
\centering
\includegraphics[width=0.9\textwidth]{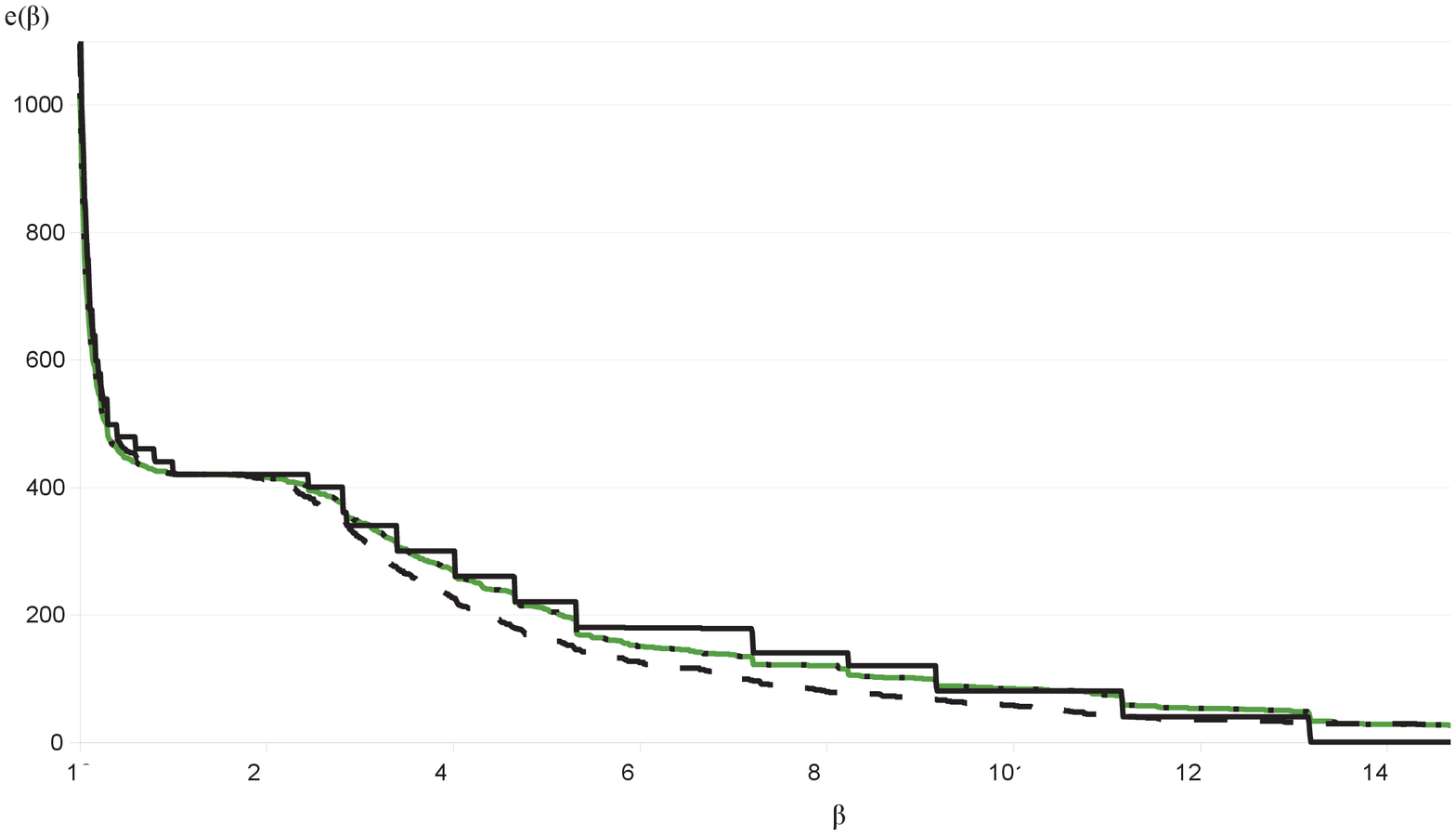}
\caption{Edge disappearance curves $e(n,\beta)$ of $\beta$-skeletons constructed over planar points arranged in 
(solid line) six nested circles with the same centre $c$ (as Fig.~\ref{circles}), 
(dashed line) six nested circles with centres slightly randomly deviated from $c$
(Fig.~\ref{circlesrandomcentres}), and (grey line with black dots) six nested circles with the same centre $c$ yet positions 
of points randomised in interval $[-2,2]$ along each axis; $n=241$, values of $\beta$ are incremented by 0.1.}
\label{circlesgraph}
\end{figure}

\begin{finding}
Presence of impurities in otherwise regular arrangements can be detected by  edge disappearance curve $e(\beta, n)$.  
\end{finding}

Let us consider a planar set where points are arranged into six nested circles all centred at the same point $c$ (Figs.~\ref{circles}a). 
When $\beta=1$ the skeleton has the following structure: every point in circle $A$ is connected by an edge to its two immediate neighbours 
in its circle, and to two neighbours in the circle included in $A$ (if there is a circle included in $A$) and two neighbours in the 
circle which includes $A$ (if there is a circle including $A$), see  Figs.~\ref{circles}b. Increase of $\beta$ from 1 to 11 leads to  
disappearance of edges connecting points in different circles (Figs.~\ref{circles}c--k). These edges disappear centrifugally. With further
increase of $\beta>11$ we observe removal of edges linking nodes in the same circles, see e.g.  (Figs.~\ref{circles}c--k).

Let us introduce a minor impurity: we make centres of circles slightly deviating, at random in a range $[-5,5]$ units along each axis, 
around centre $c$ (Figs.~\ref{circlesrandomcentres}a). With increase of $\beta$ the skeleton of such an arrangement of points loses 
majority of edges between different circles when $\beta$ reaches $7$ (Figs.~\ref{circlesrandomcentres}b--g). Few remaining edges 
are removed by $\beta=11$ (Figs.~\ref{circlesrandomcentres}h--k). Edges connecting points inside circles disappear for larger values 
of $\beta$, see e.g.  (Figs.~\ref{circlesrandomcentres}l). 

Edge disappearance curves $e(n,\beta)$, $n=241$, for the $\beta$-skeletons of the cyclic arrangements are shown in Fig.~\ref{circlesgraph}. For comparison we also added $e(n, \beta)$ for six nested circles with the same centre $c$ where position of each point is randomised in interval $[-2,2]$ 
along each axis. 

The curve $e(n,\beta)$ for circular arrangement of points with the same centre has a pronounced staircase like structure (Fig.~\ref{circlesgraph}, solid line).  The first sequence of low-height stairs is observed for $1 \leq \beta \leq 1.5$: this corresponds to removal of edges connecting points lying in 
different cycles. The second sequence of stairs, $2 \leq \beta \leq 5$ reflects removal of edges linking neighbouring points lying in the 
same cycles. Curves $e(n, \beta)$, calculated for circular arrangement of points with randomised centres and circles with randomised positions of nodes, show gradual decline in number of edges.

\section{Conclusion}

Most $\beta$-skeletons lose their edges with increase of $\beta$. The skeletons of random planar sets lose edges by power low with rate of edge disappearance proportional to a number of points in the sets. Some $\beta$-skeletons conserve their edges for any $\beta$ as large as it could be. 
These are usually skeletons built on a regularly arranged points of planar sets. We found that even minuscule impurity in the regular arrangement of points leads to propagation of edge loss wave across the otherwise stable skeleton.  This indicates that presence of random components in a planar set may lead to a higher  rate of $\beta$-driven edge disappearance. By comparing  edge disappearance curves of non-random and random planar sets (with the same number of nodes) we found that $\beta$-skeletons of random sets have larger number of edges for small values of $\beta$ (up to $\beta=2.5-3$) yet exhibit higher rate of edge loss. In examples studied skeletons of random sets lose their edges 1.5-2.5 times faster than skeletons of non-random set. For large values of $\beta$ ($\beta>25$) a number of edges in $\beta$-skeletons of non-random planar sets is over twice a number of edges in random sets. We hypothesise that by subjecting a $\beta$-skeleton of a planar set to $\beta$-driven edge removal we can discriminate between random and non-random sets. To prove the hypothesis and make the approach applicable to image classification we must collect statistics form much larger number of non-random planar sets. This will be a topic of further studies.

 \end{document}